\documentclass{article}

\usepackage{amsmath,amsfonts,epsfig,amsthm,amssymb}

\def\a{{\mathbf a}}

\def\g{{\mathbf g}}

\def\y{\mathbf{y}}
\def\x{\mathbf{x}}

\def\X{\mathbf{X}}
\def\P{\mathbf{\Phi}}

\def\e{\mathbf{e}}
\def\R{\mathbb{R}}
\def\C{\mathbb{C}}
\def\A{\mathcal{A}}

\def\H{\mathcal{H}}
\def\L{\mathcal{L}}

\newcommand{\refeq}[1] {equation (\ref{#1})}

\newtheorem{thm}{Theorem}
\newtheorem{lem}[thm]{Lemma}

\newtheorem{cor}[thm]{Corollary}

\newtheorem{mydef}{Definition}

\begin{document}

\title{Sampling and reconstructing signals from a union of linear subspaces}
%
\author{Thomas Blumensath
\thanks{Applied Mathematics, School of Mathematics, University of Southampton, University Road, Southampton, SO17 1BJ, UK}
}

\markboth{Version: \today}{}

\maketitle

\begin{abstract}
In this note we study the problem of sampling and reconstructing signals which are assumed to lie on or close to one of several subspaces of a Hilbert space. Importantly, we here consider a very general setting in which we allow infinitely many subspaces in infinite dimensional Hilbert spaces. This general approach allows us to unify many results derived recently in areas such as compressed sensing, affine rank minimisation and analog compressed sensing.  

Our main contribution is to show that a conceptually simple iterative projection algorithms is able to recover signals from a union of subspaces whenever the sampling operator satisfies a bi-Lipschitz embedding condition. Importantly, this result holds for all Hilbert spaces and unions of subspaces, as long as the sampling procedure satisfies the condition for the set of subspaces considered. In addition to recent results for finite unions of finite dimensional subspaces and infinite unions of subspaces in finite dimensional spaces, we also show that this bi-Lipschitz property can hold in an analog compressed sensing setting in which we have an infinite union of infinite dimensional subspaces living in infinite dimensional space. 
\end{abstract}
%

\section{Introduction}
\label{section:intro}

To motivate the general setting of this paper, we start with a review of the compressed sensing signal model in finite dimensions. In compressed sensing, sparse signals are considered. 
A class of $N$-dimensional signals $f$ in a Hilbert space is said to be $K$-sparse, if there is an orthonormal basis $\{\psi_i\}$, such that the $N$-dimensional vector $\x=[\langle f,\psi_i \rangle]_i $ has at most $K$ non-zero elements. More generally, if $\x_K$ is the best approximation to $\x$ with no more than $K$ non-zero elements, then if $\x-\x_K$ is small, $\x$ is said informally to be approximately $K$-sparse. 

In compressed sensing, a sparse signal is sampled by taking $M$ linear measurements $\tilde{y}_j=\langle f , \phi_j \rangle$. In matrix notation, this can be written as
\begin{equation}
\tilde{\y}=\P\x,
\end{equation}
where $\tilde{\y}$ is the vector of measurements $\langle f , \phi_j \rangle$ and where $\P$ is the matrix with entries $[\P]_{j,i} = \langle \psi_i,\phi_j \rangle$. 
In practice, the measurement process is never perfect and we have to account for measurement noise and inaccuracies. We thus assume that the measurements (or samples) are of the form
\begin{equation}
\y=\P\x + \e,
\end{equation}
where $\e$ is a measurement error.

Traditional sampling theory would predict that we require $N$ samples to be able to reconstruct $\x$ form the measurements. However, if $\x$ is $K$-sparse or approximately $K$-sparse, then we can often take less samples and still reconstruct $\x$ with near optimal precision \cite{candes05practical} \cite{donoho06compressed}. Importantly, reconstructing $\x$ from $\y$ can often be done using fast polynomial time algorithms. One of the conditions that has been shown to be sufficient for the reconstruction of $\x$ with many different fast algorithms is that the measurement process satisfies what is known as the Restricted Isometry Condition of a given order, where the order of the condition is related to the sparsity $K$. 

The Restricted Isometry Constant of order $K$ is generally defined as the smallest quantity $\delta_K$ that satisfies the condition
\begin{equation}
(1-\delta_K) \|\x\|_2^2 \leq \|\P\x\|_2^2 \leq (1+\delta_K) \|\x\|_2^2,
\end{equation}
for \emph{all} $K$ sparse vectors $\x$.

The sparse compressed sensing model defines a set of subspaces associated with the set of $K$-sparse vectors. Fixing the location of the $K$ non-zero elements in a vector $\x$ defines a $K$-dimensional subspace of $\R^N$. There are ${N\choose K}$ such $K$ dimensional subspaces, each for a different sparsity pattern. All $K$-sparse vectors, that is, all vectors with no more than $K$ non-zero elements, thus lie in the union of these ${N\choose K}$ subspaces. This interpretation of the sparse model led to the consideration of more general union of subspaces (UoS) as in \cite{lu07theory}, \cite{blumensath09sampling} and \cite{eldar08subspaces}. Such a generalization offers many advantages. For example, many types of data are known to be sparse in some representation, but also exhibit additional structure. These are so called structured sparse signals, an example of which are images, which are not only approximately sparse in the wavelet domain but also have wavelet coefficients that exhibit tree structures \cite{shapiro93embedded}, \cite{la05signal}. 
Apart from tree structured sparse models, structured sparse models include block sparse signal models \cite{baraniuk08model}, \cite{eldar09block}, \cite{blumensath07source} and the simultaneous sparse approximation problem \cite{cotter05mult}, \cite{chen06mult}, \cite{tropp06sim1}, \cite{tropp06sim2}, \cite{gribonval07atoms}. All of these models can be readily seen as UoS models.

However, the idea of UoS is applicable beyond constrained sparse models. For example, signals sparse in an over-complete dictionary \cite{rauhut07compressed}, \cite{blumensath07source}, the union of statistically independent subspaces as considered by Fletcher et. \cite{fletcher07rate} or signals sparse in an analysis frame \cite{elad07analysis} can all be understood from this general viewpoint. All of these examples were of finite unions of subspaces in finite dimensional space. But there is nothing that stops us from considering infinite dimensional spaces and infinite unions. In this case, the UoS model also incorporates signal models such as the finite rate of innovation model \cite{vetterli02sampling}, the low rank matrix approximation model \cite{recht08guaranteed} and the analog compressed sensing model \cite{eldar09from}. 

We here consider this general setting where we allow infinite unions. In this setting, we derive a conceptually simple and efficient computational strategy to solve linear inverse problems. To achieve this, we build on previous work of \cite{lu07theory} and \cite{blumensath09sampling}, where theoretical properties of UoS models were studied. Of importance are also the computational strategies previously suggested in \cite{eldar08subspaces} (where the authors studied block-sparse models) and in \cite{baraniuk08model} (where structured sparse signals were considered).

We here make the following contribution. We show that, if the sampling strategy satisfies a certain bi-Lipschitz embedding property (closely related to the Restricted Isometry Property known in compressed sensing), then, in a fixed number of iterations, a relatively simple iterative projection algorithm can compute near optimal estimates of signals that lie on, or close to, a given UoS model. These results are similar to those derived for $K$-sparse signals in \cite{blumensath08IHT} and for structured sparse models in \cite{baraniuk08model}. Our contribution here is to show that these results extend to more general UoS models (whether finite or infinite) as long as the bi-Lipschitz embedding property holds.

\subsection{Sampling and the union of subspaces models}

Union of subspaces models have been considered in \cite{lu07theory}, \cite{blumensath09sampling} and \cite{eldar08subspaces}. 
To formally define the UoS model in a general Hilbert space $\H$, consider a set of arbitrary subspaces $\A_i\subset\H$. We then define the UoS as the set 
\begin{equation}
\A=\bigcup \A_i.
\end{equation}


In analogy with compressed sensing, sampling of an element $\x\in\H$ is done using a linear operator $\P : \H \rightarrow \L$, where $\L$ is some Hilbert space. We then write the observations as
\begin{equation}
\y=\P\x + \e,
\end{equation}
where $\e\in \L$ is again an error term.

\subsection{The bi-Lipschitz condition}
In order to guarantee stability, it is necessary to impose a bi-Lipschitz condition on $\P$ as a map from $\A$ to $\L$. 
\begin{mydef}
We say that $\P$ is bi-Lipschitz on a set $\A$, if there exist constants $0<\alpha\leq\beta$, such that for all $\x_1, \x_2\in\A$ 
\begin{equation}
\alpha \|\x_1+\x_2\|^2\leq  \|\P(\x_1+\x_2)\|^2\leq\beta \|\x_1+\x_2\|^2.
\end{equation}
The bi-Lipschitz constants of $\P$ on $\A$ are the largest $\alpha$ and smallest $\beta$  for which the above inequalities hold for all $\x_1, \x_2\in\A$.
\end{mydef}
Whilst $\beta$ is the square of the Lipschitz constant of the map $\P$ (as a map from $\A$ to $\L$), $1/\alpha$ is the square of the Lipschitz constant of the inverse of $\P$ defined as a map from $\P\A\subset \L$ to $\A$. Note that the requirement $\alpha>0$ is equivalent to the requirement that $\P$ is one to one as a map from $\A$ to $\L$. Therefore, the inverse of $P$ is well defined as a function from $\P\A$ to the set $\A$ whenever $\alpha>0$.

\subsection{Proximal sets and projections}
\label{subsection:sep}

When dealing with infinite dimensions and infinite unions, extra care has to be taken. In order to guarantee the existence of (possibly non-unique) best approximations of elements in $\H$ with elements from $\A$, additional assumptions on $\A$ are required. In addition to the assumption that $\A$ is a closed set, we assume that the set $\A$ is \emph{proximal}, that is, that for all $\x\in \H$ the set 
\begin{equation}
p_\A(\x)=\{\tilde\x:\tilde\x\in\A, \|\tilde\x-\x\|=\inf_{\hat{\x}\in\A} \|\hat\x-\x\|\}
\end{equation}
is non-empty.
For proximal sets $\A$ we can therefore define a projection as any point $\x_\A$ that satisfies
\begin{equation}
\|\x-\x_\A\|=\inf_{\hat{\x}\in\A} \|\hat\x-\x\|.
\end{equation}
Note that $\x_\A$ is the orthogonal projection of $\x$ onto one of the subspaces $\A_i$. We write this projection as
\begin{equation}
P_{\A}(\x)=S(p_\A(\x));
\end{equation} 
where $S$ is a set valued operator that returns a single element of the set $p_\A(\tilde\x)$. How this element is chosen in practice does not influence the theoretical results derived here so that we do not specify any particular approach in this paper.


\section{The optimal solution}

In order to talk about optimal solutions, we require the existence of a projection of a point $\y\in\L$ onto the set $\P\L$. Note that we assume $\A$ to be closed which implies that $\P\A$ is closed if $\P$ is be-Lipschitz. However, as stated above, closedness of $\P\A$ is not sufficient to show that the projection onto $\P\A$ exists. In this section we therefore also assume that $\P\A$ is \emph{proximal}. 

More formally, consider
\begin{equation}
\inf_{\tilde\x\in\A}\|\y-\P\tilde\x\|.
\end{equation}

As $\P\A$ is assumed to be \emph{proximal}, we can define optimal solutions as those elements $\x_{opt}\in\A$ for which 
\begin{equation}
\|\y-\P\x_{opt}\| = \inf_{\tilde\x\in\A}\|\y-\P\tilde\x\|.
\end{equation}

Alternatively, instead of considering \emph{proximal} sets $\P\A$, we could define $\epsilon$ optimal points as those points $\x_{opt}^{\epsilon}\in\A$ for which
\begin{equation}
\|\y-\P\x_{opt}\| \leq \inf_{\tilde\x\in\A}\|\y-\P\tilde\x\|+\epsilon.
\end{equation}
The results derived below then still hold but will include additional $\epsilon$ terms. To avoid carrying around these additional terms, we here assume that $\P\A$ is a \emph{proximal} subset of $\L$.

The bi-Lipschitz condition guarantees that $\P$ is one to one as a function from $\A$ to $\L$, that is, it maps distinct points form $\A$ into distinct points in $\L$. We are therefore able, at least in theory, to invert $\P$ on $\A$. The condition also guarantees stability in that, for any $\x\in\A$, if we are given an observation $\y=\P\x+\P\hat\x+\e$, where $\e\in\L$ and $\hat\x\in\H$ are general errors, then we could, at least in theory, recover a good approximation of $\x$ as follows. We let $\hat\y$ be the projection of $\y$ onto the closest element in $\P\A$. We then look for the unique $\x\in\A$ for which $\hat\y=\P\hat\x$. As will be shown more rigorous below, the bi-Lipschitz property of $\P$ then guarantees that $\hat\x$ is close to $\x$.

We now show that all $\x_{opt}$ are basically optimal if the bi-Lipschitz property holds, that is, we can't define an estimate that performs substantially better.
  
Let us first derive an upper bound for the error. Note that by definition of $\x_{opt}$, $\|\y-\P\x_{opt}\|\leq \|\y-\P\x_\A\|$, where we define $\x_\A = P_{\A}(\x)$. Defining $\e_{opt}=\y-\P\x_{opt}$ and $\e_\A=\y-\P\x_\A$ we thus have
\begin{eqnarray}
\|\x-\x_{opt}\|
&   \leq  & \|\x_\A-\x_{opt}\| +\|\x-\x_\A\| \nonumber \\
& \leq & \frac{1}{\sqrt{\alpha}}\|\P(\x_\A-\x_{opt})\| + \|\x-\x_\A\| \nonumber \\
& = & \frac{1}{\sqrt{\alpha}}\|\e_\A - \hat{\e}\| + \|\x-\x_\A\| \nonumber \\
& = & \frac{1}{\sqrt{\alpha}}\|\e_\A\| + \frac{1}{\sqrt{\alpha}}\| \hat{\e}\|  + \|\x-\x_\A\|  \nonumber \\
& \leq & \frac{2}{\sqrt{\alpha}}\|\e_\A\|  + \|\x-\x_\A\|  \nonumber,
\end{eqnarray}
where the second inequality is due to the Lipschitz property and the last inequality due to the fact that $\|\e_{opt}\|\leq\|\e_\A\|$.

We furthermore have the following 'worst case' lower bound
\begin{thm}
For each $\x$ there exists an $\e$, such that  
\begin{equation}
\|\x-\x_{opt}\| \geq   \sqrt{\frac{0.5}{\beta}} \|\e_\A\|   + \|\x-\x_\A\|  \nonumber \\
\end{equation}
\end{thm}
\begin{proof}
We have the lower bound
\begin{eqnarray}
\|\x-\x_{opt}\|^2 
&   =  & \|\x_\A-\x_{opt}\|^2 +\|\x-\x_\A\|^2 -2\langle (\x_\A-\x_{opt}),(\x-\x_\A) \rangle\nonumber \\
& \geq & \frac{1}{\beta}\|\P(\x_\A-\x_{opt})\|^2 + \|\x-\x_\A\|^2 -2\langle (\x_\A-\x_{opt}),(\x-\x_\A)  \rangle, \nonumber 
\end{eqnarray}
where from now on we simplify the notation and write $\langle \cdot,\cdot\rangle$ for the real part of the inner product $Re\langle \cdot,\cdot\rangle$.

Let $\mathcal{C}_i$ be the cone of elements $\y \in \L$ for which $\x_{opt}\in\A_i$. Because $\x_\A$ is the orthogonal projection of $\x$ onto the closest subspace, if $\x\in\A_i$, then $\x-\x_\A$ is orthogonal to $\A_i$.
Thus, if $\x_\A\in\A_i$ and if $\y\in\mathcal{C}_i$, then
\begin{equation}
\langle (\x_\A-\x_{opt}),(\x-\x_\A)  \rangle = 0.
\end{equation}

Also, for all $\y\in\mathcal{C}_i$, because $\e_{opt}=\y-\P\x_{opt}$ is orthogonal to $\P\x_\A-\P\x_{opt}$, 
\begin{equation}
 \frac{1}{\beta}\|\P(\x_\A-\x_{opt})\|^2 + \frac{1}{\beta} \|\e_{opt}\|^2 =  \frac{1}{\beta}\|\e_\A\|^2,
\end{equation}
so that for all $\y\in\mathcal{C}_i$
\begin{eqnarray}
\|\x-\x_{opt}\|^2 & \geq & \frac{1}{\beta}\|\e_\A\|^2 - \frac{1}{\beta}  \|\e_{opt}\|^2 + \|\x-\x_\A\|^2.  \nonumber 
\end{eqnarray}
We can now choose $\e=c\P\x_\A$, where $c>1$ is chosen large enough for $\y$ to be in $\mathcal{C}_i$. Because $\e_{opt}$ is orthogonal to $\P\A_i$, $\|\e_{opt}\|$ is constant as a function of $c$, whilst $\|\e_\A\|$ increases for $c>1$. We can thus choose $c$ (and thus $\e$) such that $\y\in\mathcal{C}_i$ and 
\begin{equation}
-\|\e_{opt}\|^2 > - 0.5\|\e_\A\|^2 +  \sqrt{2\beta}\|\e_\A\|\|\x-\x_\A\|,
\end{equation}
so that for all $\x$ there is an $\e$ such that 
\begin{eqnarray}
\|\x-\x_{opt}\|^2 & \geq & 2 \frac{0.5}{\beta}\|\e_\A\|^2 + \|\x-\x_\A\|^2 +\sqrt{\frac{0.5}{\beta}}\|\e_\A\|\|\x-\x_\A\|,  \nonumber 
\end{eqnarray}
from which the theorem follows.
\end{proof}

\section{The Iterative Projection Algorithm}
\label{section:IPA}

Calculating $\x_{opt}$ is highly non-trivial for most $\P$ and $\A$. We therefore propose an iterative algorithm and show that under certain conditions on $\alpha$ and $\beta$ we can efficiently calculate solutions whose error is of the same order as that achieved by $\x_{opt}$. In order for our algorithm to be applicable, we require that we are able to efficiently calculate the projection of any $\x\in\H$ onto the closest $\A_i$ (which therefore has to be well defined). 

The Iterative Projection Algorithm is a generalization of the Iterative Hard Thresholding algorithm of \cite{kingsbury03iterative}, \cite{blumensath08thresh} and \cite{blumensath08IHT} to general UoS models.

Assume $\A$ is \emph{proximal}. Given $\y$ and $\P$, let $\x^{0}=\mathbf{0}$. The Iterative Projection Algorithm is the iterative procedure defined by the recursion
\begin{equation}
\x^{n+1} = P_{\A}(\x^{n} + \mu \P^T(\y-\P\x^{n} )),
\end{equation}
where the non-linear operator $P_{\A}(\a)$ is defined in subsection \ref{subsection:sep}.  

In many problems, calculation of $P_{\A}(\a)$ is much easier than a brute force search for $\x_{opt}$. For example, in the $K$-sparse model, $P_{\A}(\a)$ simply keeps the largest (in magnitude) $K$ elements of $\a$ and sets the other elements to zero, whilst in the low rank matrix approximation problem, different efficient projections have been defined in \cite{recht08guaranteed}. Furthermore, the above algorithm only requires the application of $\P$ and its adjoint, which can often be computed efficiently. Importantly, the next result shows that under certain conditions, not only does the algorithm calculate near optimal solutions, it does so in a fixed number of iterations (depending only on a form of signal to noise ratio)! 

We have the following main result.
\begin{thm}
\label{thm:main}
Let $\A$ be a \emph{proximal} subset of $\H$. Given $\y=\P\x+\e$ where $\x$ is arbitrary. 
Assume $\P$ is bi-Lipschitz as a map from $\A$ to $\L$ with constants $\alpha$ and $\beta$.
If $\beta\leq \frac{1}{\mu}<1.5\alpha$, then, after
\begin{equation}
n^\star = \left\lceil 2 \frac{\ln(\delta\frac{\|\e_\A\|}{\|\x_\A\|})}{\ln(2/(\mu\alpha)-2)} \right\rceil
\end{equation}
iterations, the Iterative Projection Algorithm calculates a solution $\x^{n^\star}$ satisfying
\begin{equation}
\|\x-\x^{n^\star}\| \leq (c^{0.5}+\delta)\|\e_\A\| + \|\x_{\A}-\x\|,
\end{equation}
where $c \leq \frac{4}{3\alpha-2\mu}$ and $\tilde \e = \P(\x-\x_{\A})+e$.
\end{thm}

Note that this bound is of the same order as that derived for $\x_{opt}$.

The above theorem has been proved for the $K$-sparse model in \cite{blumensath08IHT} and for constraint sparse models in \cite{baraniuk08model}. Our main contribution is to show that it holds for general UoS constrained inverse problems\footnote{It might also be worth noting that the proof of Theorem \ref{thm:main} is not only valid for union of subspaces, but holds for arbitrary subsets of $\A\subset\H$ for which $\P$ satisfies the bi-Lipschitz requirement. A more detailed discussion of this fact is left for an upcoming publication.}, as long as the bi-Lipschitz property holds with appropriate constants. 

To derive the result, we pursue a slightly different approach to that in \cite{blumensath08IHT} and \cite{baraniuk08model} and instead follow the ideas of \cite{garg09grad}. The proof is based on the following lemma.
\begin{lem}
\label{lem:ProjOpt}
If $\frac{1}{\mu}\geq\beta$ then, using $\x^{n+1}=P_{\A}(\x^n+\mu\P^*(\y-\P\x^n))$, we have
\begin{eqnarray}
& &\|\y-\P\x^{n+1}\|^2 - \|\y-\P\x^n\|^2 \nonumber \\
& \leq& -\langle(\x_\A-\x^n), \g\rangle + \frac{1}{\mu}\|\x_\A-\x^n \|^2,
\end{eqnarray}
where $\g=2\P^*(\y-\P\x^n)$.
\end{lem}
\begin{proof}
The left hand side in the equality of the lemma can be bounded by
\begin{eqnarray}
& 		&\|\y-\P\x^{n+1}\|^2 - \|\y-\P\x^n\|^2 \nonumber \\
&  =  & -\langle(\x^{n+1}-\x^n), \g\rangle + \|\P(\x^{n+1}-\x^n) \|^2 \nonumber \\
&\leq & -\langle(\x^{n+1}-\x^n), \g\rangle + \frac{1}{\mu}\|(\x^{n+1}-\x^n) \|^2 \nonumber
\end{eqnarray}

We will now show that $\x^{n+1}=H_\A(\x^n+\frac{\mu}{2}\g)$ minimizes $-\langle(\tilde\x-\x^n), \g\rangle + \frac{1}{\mu}\|(\tilde\x-\x^n) \|^2$ over all $\tilde\x\in\A$ so that $\x_\A\in\A$ implies that
\begin{equation}
 -\langle(\x^{n+1}-\x^n),\g\rangle + \frac{1}{\mu}\|(\x^{n+1}-\x^n) \|^2
 \leq-\langle(\x_\A-\x^n),\g\rangle + \frac{1}{\mu}\|(\x_\A-\x^n) \|^2, 
 \end{equation} 
from which the lemma will follow.

We write the infimum of $-\langle(\tilde\x-\x^n), \g\rangle + \frac{1}{\mu}\|(\tilde\x-\x^n) \|^2$ as
\begin{eqnarray}
&   & \inf_{\x\in\A} (- \langle\x,\g\rangle + \langle\x^{n},\g\rangle + \frac{1}{\mu}\|(\x-\x^n) \|^2) \nonumber \\
& \propto & \inf_{\x\in\A} (- \mu \langle\x,\g\rangle + \langle \x,\x\rangle + \|\x^n\|^2 -2\langle \x,\x^{n} \rangle) \nonumber \\
& \propto & \inf_{\x\in\A}  (- \mu \langle\x,\g\rangle +  \langle \x,\x\rangle  -2\langle \x,\x^{n} \rangle) \nonumber \\
& \propto & \inf_{\x\in\A} \|\x-\x^n-\frac{\mu}{2}\g\|^2 \nonumber \\
& = &\|\x^{n+1}-\x^n-\mu\P^*(\y-\P\x^n)\|^2,\nonumber
\end{eqnarray}
where the last equality comes from the definition of  $\x^{n+1}=P_{\A}(\x^n+\mu\P^*(\y-\P\x^n))$. Thus, the infimum of $-\langle(\tilde\x-\x^n), \g\rangle + \frac{1}{\mu}\|(\tilde\x-\x^n) \|^2$ is proportional to $\inf_{\x\in\A} \|\x-\x^n-\frac{\mu}{2}\g\|^2$ so that $\x^{n+1}$ simultaneously minimises both quantities.

\end{proof}

\begin{proof}[Proof of Theorem \ref{thm:main}]
Let $\x_{\A}=P_{\A}(\x)$, so that the triangle inequality implies that.
\begin{equation}
\|\x-\x^{n+1}\|\leq \|\x_{\A}-\x^{n+1}\| + \|\x_{\A}-\x\|.
\end{equation}
The square of the first term on the right is bounded using the bi-Lipschitz property of $\P$
\begin{eqnarray}
\|\x_{\A}-\x^{n+1}\|^2 
&\leq& \frac{1}{\alpha}  \|\P(\x_{\A}-\x^{n+1})\|^2. 
\end{eqnarray}
We expand this, so that
\begin{eqnarray}
& &\|\P(\x_\A-\x^{n+1}) \|^2 = \|\y -\P\x^{n+1} -\e_\A \|^2 \nonumber \\ 
& \leq &\|\y-\P\x^{n+1}\|^2 + \|\e_\A\|^2 -2\langle\e_\A,(\y-\P\x^{n+1})\rangle \nonumber \\ 
&\leq& \|\y-\P\x^{n+1}\|^2 +\|\e_\A\|^2 + \|\e_\A\|^2 +\|\y-\P\x^{n+1}\|^2 \nonumber \\
& = & 2 \|\y-\P\x^{n+1}\|^2 +  2 \|\e_\A\|^2,
\end{eqnarray}
where the last inequality follows from $-2\langle\e_\A,(\y-\P\x^{n+1})\rangle\leq \|\e_\A\|\|(\y-\P\x^{n+1})\|\leq 0.5(\|\e_\A\|^2+\|(\y-\P\x^{n+1})\|^2)$. 

We will now show that under the Lipschitz assumption of the theorem, the first term on the right is bounded by 
\begin{equation}
\|\y-\P\x^{n+1}\|^2\leq (\mu-\alpha)\|(\x_\A-\x^n)\|^2 + \|\e_\A\|^2.
\end{equation}
To show this, we write
\begin{eqnarray}
& &  \| \y-\P\x^{n+1}\|^2 - \|\y-\P\x^n\|^2 \nonumber \\
&\leq & -2\langle(\x_\A-\x^n),\P^*(\y-\P\x^n)\rangle + \frac{1}{\mu}\|\x_\A-\x^n \|^2 \nonumber \\
& =   & -2\langle(\x_\A-\x^n),\P^*(\y-\P\x^n)\rangle + \alpha \|\x_\A-\x^n \|^2 +(\frac{1}{\mu}-\alpha)\|\x_\A-\x^n \|^2 \nonumber \\ 
& \leq   & -2\langle(\x_\A-\x^n),\P^*(\y-\P\x^n)\rangle + \|\P(\x_\A-\x^n) \|^2 +(\frac{1}{\mu}-\alpha)\|\x_\A-\x^n \|^2 \nonumber \\ 
&   =  & \|\y-\P\x_\A\|^2-\|\y-\P\x^{n}\|^2 +(\frac{1}{\mu}-\alpha)\|\x_\A-\x^n \|^2 \nonumber \\ 
& = & \|\e_\A\|^2-\|\y-\P\x^{n}\|^2 + (\frac{1}{\mu}-\alpha)\|(\x_\A-\x^n) \|^2 
\end{eqnarray}
where the first inequality is due to Lemma \ref{lem:ProjOpt}.

We have thus shown that
\begin{equation}
\|\x_{\A}-\x^{n+1}\|^2 
\leq 2\left( \frac{1}{\mu\alpha}-1 \right)\|(\x_\A-\x^n) \|^2 + \frac{4}{\alpha}\|\e_\A\|^2. 
\end{equation}

Under the condition of the Theorem, $2 (\frac{1}{\mu\alpha}-1)<1$, so that we can iterate the above expression
\begin{equation}
\|\x_{\A}-\x^{k}\|^2 
\leq \left(2\left(\frac{1}{\mu\alpha}-1\right)\right)^k\|\x_\A \|^2 + c\|\e_\A\|^2, 
\end{equation}
where $c \leq \frac{4}{3\alpha-2\frac{1}{\mu}}$.

In conclusion, we have
\begin{eqnarray}
\|\x-\x^{k}\| 
&\leq& \sqrt{\left(2\frac{1}{\mu\alpha}-2\right)^k\|\x_\A \|^2 + c\|\e_\A\|^2} + \|\x_{\A}-\x\| \nonumber \\
&\leq& \left(2\frac{1}{\mu\alpha}-2\right)^{k/2} \|\x_\A \| + c^{0.5}\|\e_\A\| + \|\x_{\A}-\x\|, 
\end{eqnarray}
which means that after $k^\star = \left\lceil 2 \frac{\ln(\delta\frac{\|\e_\A\|}{\|\x_\A\|})}{\ln(2/(\mu\alpha)-2)} \right\rceil$ iterations we have
\begin{equation}
\|\x-\x^{k^\star}\| \leq (c^{0.5}+\delta)\|\e_\A\| + \|\x_{\A}-\x\|.
\end{equation}


\end{proof}

\subsection{A remark on $\e_\A$}
For readers familiar with the literature on compressed sensing a remark is in order. In our general result, we have written the bound on the result in terms of $\|\e_\A\| =\|\P(\x-\x_\A)\e\|$. This is the most general statement in which we do not assume additional structure on $x-x_\A$ and $\P$. This differs from results in sparse inverse problems, where, under the bi-Lipschitz property, $\e_\A$ is proportional to $\|\x-\x_K\| + \frac{\|\x-\x_K\|_1}{K}$. Here $x_K$ is the best $K$-term approximation to $\x$. It also differs from results derived in \cite{baraniuk08model} where $\A$ satisfies certain nesting properties and where a Restricted Amplification property is used to bound $\|\e_\A\|$ by a function of $\x-\x_\A$. Unfortunately, in the general setting of this paper, such a bound is not possible without additional assumptions and the best one could hope for would be to bound $\|\e_\A\|$ by $\|\P\|\|\x-\x_\A\|+\|\e\|$.


\section{Examples of bi-Lipschitz embeddings}

The bi-Lipschitz property depends on both, $\P$ and $\A$. In this section we will study three particular cases from the literature. For the first two cases, bi-Lipschitz maps have already been studied and we here review the main results before deriving a new result that demonstrates how such properties can be proved even in an infinite dimensional setting.

\subsection{Finite Unions of Finite dimensional Subspaces}

We start with the finite dimensional setting and with unions of finite dimensional subspaces. In particular, let $\A$ be the union of $L<\infty$ subspaces each of dimension no more than $K$ and let $\P\A\subset\R^M$.
This is an important special case of UoS models which covers many of the problems studied in practice, such as the $K$-sparse models used in compressed sensing \cite{candes05practical}, \cite{donoho06compressed}, block sparse signal models \cite{baraniuk08model}, \cite{eldar09block}, \cite{blumensath07source}, the simultaneous sparse approximation problem \cite{cotter05mult}, \cite{chen06mult}, \cite{tropp06sim1}, \cite{tropp06sim2}, \cite{gribonval07atoms}, signals sparse in an over-complete dictionary \cite{rauhut07compressed}, \cite{blumensath07source}, the union of statistically independent subspaces as considered by Fletcher et. \cite{fletcher07rate} and signals sparse in an analysis frame \cite{elad07analysis}. 
Finite unions of finite dimensional subspaces have therefore been studied in, for example, \cite{blumensath09sampling} where the following result was derived.
\begin{thm}
\label{theorem:ind2}
For any $t>0$, let
\begin{equation}
M\geq \frac{2}{c \delta_{\mathcal{A}}}  \left(\ln (2{L})  + {2K} \ln\left(\frac{12}{ \delta_{\mathcal{A}}}\right) + t\right),
\end{equation}
then there exist a $\P$ and a constant $c>0$ such that
\begin{equation}
\label{equation:inds}
(1-\delta_{\mathcal{A}} (\P))\|\y_1-\y_2\|_2^2 \leq \|\P(\y_1-\y_2)\|_2^2 \leq (1+\delta_{\mathcal{A}} (\P)) \|\y_1-\y_2\|_2^2
\end{equation}
holds for all $\y_1,\y_2$ from the union of ${L}$ arbitrary $K$ dimensional subspaces $\mathcal{A}$.
What is more, if $\P$ is an $M\times N$ matrix generated by randomly drawing i.i.d. entries from an appropriately scaled subgaussian distribution\footnote{Examples of these distributions include the Gaussian distribution and random variables that are $\pm \frac{1}{\sqrt{N}}$ with equal probability \cite{baraniuk06johnson}  \cite{rauhut07compressed}.}, then this matrix satisfies \refeq{equation:inds} with probability at least
\begin{equation}
1- e^{-t}.
\end{equation}
The constant $c$ then only depends on the distribution of the entries in $\P$ and is $c=\frac{7}{18}$ if the entries of $\P$ are i.i.d. normal.
\end{thm}

\subsection{Infinite Unions of Finite dimensional Subspaces in $\R^N$}
 
Recently, similar results could also be derived for a union of infinitely many subspaces. In \cite{recht08guaranteed} minimum rank constrained linear matrix valued inverse problems are studied. These problems are another instance of the linear inverse problem studied in this paper and can be stated as follows: Find a matrix $\X\in\R^{m\times n}$ with rank no more than $K$, such that $\y=P(\X)$, where $P(\cdot)$ is a linear function that maps $\R^{m\times n}$ into $\R^M$. Vectorising $\X$ as an element $\x\in\R^N$ and by writing $P$ in matrix form, we have the linear inverse problem where $\A$ is the set of vectorised matrices with rank at most $K$. This problem was solved with the Iterative Projection Algorithm in \cite{goldfarb09conv} where it was also shown that 
\begin{thm}
If $P$ is a random \emph{nearly isometrically distributed} linear map\footnote{See \cite{goldfarb09conv} for an exact definition of nearly isometrically distributed linear maps. An example would again be if the matrix $\P$ has appropriately scaled i.i.d. Gaussian entries.}, then with probability $1-e^{-c_1N}$,
\begin{equation}
(1-\delta)\|\X_1-\X_2\|_F\leq\|P(\X_1-\X_2)\|\leq(1+\delta)\|\X_1-\X_2\|_F
\end{equation} 
for all rank $K$ matrices $\X_1\in\R^m\times n$ and $\X_2\in\R^m\times n$, whenever $N\geq c_0 K(m+n)log (mn)$, where $c_1$ and $c_0$ are constants depending on $\delta$ only.
\end{thm}
 
\subsection{Infinite Unions of Infinite dimensional Subspaces}

We now show that non-trivial bi-Lipschitz embeddings also exist between infinite dimensional spaces $\H$ and $\L$, where $\A$ is an infinite union of infinite dimensional subspaces in $\H$. We here consider the example from \cite{mishali09blind}. A continuous real valued time series $x(t)$ is assumed to be band-limited, that is, its Fourier transform $\mathcal{X}(f)$ is assumed to be zero apart from the set $S\subset[-B_N\ B_N]$. Furthermore, the support of $\mathcal{X}(f)$ is assumed to be 'sparse' in the sense that we can write $S$ as the union of $K$ intervals of 'small' bandwidth $B_K$, i.e. $S \subset \bigcup_{k=1}^K [d_k \ d_k+B_K]$, where the $d_k$ are arbitrary scalars from the interval $[0\ B_N-B_K]$. Note, due to symmetry, we only consider the support in the positive interval $[0\ B_N]$. Crucially, we assume that $K B_K<B_N$, so that $\mathcal{X}(f)$ is zero for most (in terms of Lebesgue measure) $f$ in $[0\  B_N]$. Fixing the support $S$, $\mathcal{X}(f)$ and therefore $x(t)$ lie on a subspace of the space of all square integrable functions with bandwidth $B_N$. If $K B_K<B_N$, then there are infinitely many distinct sets $S$ satisfying this definition, so that $x(t)$ lies in the union of infinitely many infinite dimensional subspaces. 

Classical sampling theory tells us that there exists sampling operators that map band-limited functions into $\ell_2$. What is more, these sampling operator are not only one to one, but also isometric, that is, bi-Lipschitz embeddings with $\alpha=1$ and $\beta=1$. These sampling operators are given by the Nyquist sampling theorem, which only takes account of the bandwidth $B_N$, but does not consider additional structure in $\A$. To improve on the classical theory, we are thus interested in sampling schemes with a sampling rate that is less than the Shannon rate.

To this end, we show that there exist bi-Lipschitz embeddings of functions from $\A$ into the space of band-limited signals with bandwidth $B_M$, where $B_M<B_N$. Combining this embedding with the standard (isometric) Nyquist sampling kernel for functions with bandwidth $B_M$, gives a stable sampling scheme where the sampling rate is $2B_M$ instead of $2B_N$. The iterative projection algorithm will therefore also be applicable to this sampling problem. It is worth noting that the bi-Lipschitz embedding property shown here not only guarantees invertability of the sampling process, which was demonstrated for the problem under consideration in \cite{mishali09blind}, but also guarantees stability of this inverse.

Our treatment here is theoretical in nature and is meant as an example to show how bi-Lipschitz embeddings can be constructed in the infinite dimensional setting, it is not meant as a fully fledged practical sampling method and many practical issues remain to be addressed.

Compressed Sensing theory has shown that there is a constant c such that there are matrices $\P\in\R^{M\times N}$ with $M\leq c K ln(N/k)$ which are bi-Lipschitz embeddings from the set of all $K$ sparse vectors in $\R^N$ to $\R^M$ \cite{candes06near}. Therefore, assume $\P$ satisfies 
\begin{equation}
\alpha \|\x\|^2\leq\|\P\x\|^2\leq \beta \|\x\|^2
\end{equation}
for all vectors $\x\in\C^N$ with no-more than $2K$ non-zero elements. 

The following sampling approach is basically that proposed in \cite{eldar09from} and is based on mixing of the spectrum of $x(t)$. It uses a matrix $\P\in\R^{M\times N}$ to define this mixing procedure. Our contribution is to show that if the matrix $\P$ has the bi-Lipschitz property with constants $\alpha$ and $\beta$, then so will this sampling operator.

Let $\A\subset L^2_\C([0\ B_N])$ be the subset of the set of square integrable real valued functions whose Fourier transform has positive support $S\subset [0\ B_N]$, where $S$ is the union of no more than $K$ intervals of width no more than $B_K$.
Let $M=\left\lceil B_N/B_K \right\rceil$ and let $B_M=M B_K$. We then split the interval $[0\ B_N]$ into $M$ blocks of length $B_K$ as follows.
Let $S_j$ be the interval $[(j-1)B_K\ jB_K)$ for integers $1\leq j \leq N-1$ and $S_N = [(j-1)B_K\ B_N]$. Similarly, let $\tilde{S}_i$ be the interval $[(i-1)B_K\ iB_K)$ for integers $1\leq i \leq M-1$ and $S_M = [(i-1)B_K\ iB_K]$.
We can then define a linear map from $x(t)$ to $y(t)$ by mapping the Fourier transform of $x(t)$ into the Fourier transform of $y(t)$ as follows
\begin{equation}
\label{equation:infmap}
\mathcal{Y}(S_i)=\sum_{j=1}^N [\P]_{i,j} \mathcal{X}(S_j),
\end{equation}
where we use the convention that $\mathcal{X}(f)=0$ for $f>B_N$. In words, the new function has the Fourier transform $\mathcal{Y}$ (defined by symmetry also for $f<0$) which is constructed by concatenating $M$ functions of length $B_K$. Each of these blocks is a weighted sum of the N blocks of $\mathcal{X}$, where the weights are the entries of the matrix $\P$.

We have the following result
\begin{thm}
Let $\A\subset L^2_\C([0\ B_N])$ be the subset of the set of square integrable real valued functions whose Fourier transform has positive support $S\subset [0\ B_N]$, where $S$ is the union of no more than $K$ intervals of width no more than $B_K$.
If the matrix $\P\in\R^{M\times N}$ is bi-Lipschitz as a map from the set of all $K$-sparse vectors in $\R^N$ into $\R^M$, with bi-Lipschitz constants $\alpha$ and $\beta$, then the map defined by equation \ref{equation:infmap} is a bi-Lipschitz map from $\A$ to $L^2_\C([0\ B_M])$ such that
\begin{equation}
\alpha\|\mathcal{X}_1-\mathcal{X}_2\|^2 \leq   \|\mathcal{Y}_1-\mathcal{Y}_2\|^2_2 \leq  \beta\|\mathcal{X}_1-\mathcal{X}_2\|^2_2  ,
\end{equation}
for all $\mathcal{X}_1,\mathcal{X}_2 \in\A$.
\end{thm}

\begin{proof}
To see that this map is bi-Lipschitz from $\A$ to $L^2([0\ B_M])$, consider stacking up the blocks $\mathcal{Y}(\tilde{S}_i)$ and $\mathcal{X}(S_j)$ in two vectors.
For $f\in[0\ B_K)$ we use $f_i=(i-1)*B_K+f$ and $\tilde{f}_j=(j-1)*B_K+f$ and define the vectors
\begin{equation}
\y(f)=
\left[ \begin{array}{cccc}
\mathcal{Y}(\tilde{f}_1) \\
\mathcal{Y}(\tilde{f}_2) \\
\cdots \\
\mathcal{Y}(\tilde{f}_M) \end{array} \right]
=
\P 
\left[ \begin{array}{cccc}
\mathcal{X}(f_1) \\
\mathcal{X}(f_2) \\
\cdots \\
\mathcal{X}(f_N) \end{array} \right] = \P\x(f).
\end{equation}
This model is known as an infinite measurement vector model \cite{eldar09from}.

Using the norm of $L^2$, we can write
\begin{eqnarray}
\|\mathcal{Y}_1-\mathcal{Y}_2\|^2 = \int_0^{B_M}{ (\mathcal{Y}_1(f)-\mathcal{Y}_2(f))^2  \ df} \nonumber \\
=  \int_0^{B_K}{\sum_i (\mathcal{Y}_1((i-1)B_K+f)-\mathcal{Y}_2((i-1)B_K+f))^2  \ df} \nonumber \\
=  \int_0^{B_K}{\|\y_1(f)-\y_2(f)\|^2_2  \ df} \nonumber \\
=  \int_0^{B_K}{ \|\P\x_1(f)-\P\x_2(f)\|^2_2  \ df}.
\end{eqnarray}
Noting that for fixed $f$, the vectors $\x_1(f)$ and $\x_2(f)$ are $K$-sparse, the bi-Lipschitz property of $\P$ leads to the inequalities
\begin{equation}
 \int_0^{B_K}{\alpha\|\x_1(f)-\x_2(f)\|^2_2  \ df} \leq   \int_0^{B_K}{\|\P\x_1(f)-\P\x_2(f)\|^2_2  \ df} \leq  \int_0^{B_K}{\beta\|\x_1(f)-\x_2(f)\|^2_2  \ df} ,
\end{equation}
so that
\begin{equation}
\alpha\|\mathcal{X}_1-\mathcal{X}_2\|^2 \leq   \|\mathcal{Y}_1-\mathcal{Y}_2\|^2_2 \leq  \beta\|\mathcal{X}_1-\mathcal{X}_2\|^2_2  ,
\end{equation}
i.e the mapping defined above satisfies the bi-Lipschitz condition with constants $\alpha$ and $\beta$ defined by the bi-Lipschitz constants of the matrix $\P$.
\end{proof}

If we consider signals whose Fourier transform has support $S$ and if we let $|S|$ be the size of the support, then, if we assume that the support is the union of finitely many intervals of length $B_K$, then we have $|S|=K B_K$ for some $K$. If we then use $N=B_N/B_K$ and select $M$ and $B_M$ such that $B_M=M B_K$, then the fact that there are bi-Lipschitz matrices with $M=c K \ln{N/K}$ together with the above theorem implies the following corollary
\begin{cor}
Let $\A\subset L^2_\C([0\ B_N])$ be the subset of the set of square integrable real valued functions whose Fourier transform has positive support $S\subset [0\ B_N]$, where $|S|$ is bounded and where $S$ is the union of finitely many intervals of finite width.
There exist bi-Lipschitz embeddings from $\A$ to $L^2_\C([0\ B_M])$ whenever
\begin{equation}
B_M \geq c |S| \ln\left(\frac{B_N}{|S|}\right),
\end{equation}
where c is some constant.
\end{cor}  

\section{Conclusion}
We have here presented a unified framework that allows us to sample and reconstruct signals that lie on or close to the union of subspaces. The bi-Lipschitz property is necessary to guarantee stable reconstruction. We have shown that bounds on the bi-Lipschitz constants $\alpha$ and $\beta$ are sufficient for the near optimal reconstruction with the iterative projection algorithm. Whilst we have here concentrated on the general theory for arbitrary union of subspaces models, we have highlighted several more concrete examples from the literature. We could also show that bandlimited signals with 'sparse' frequency support admit sub-Nyquist sampling methods that are bi-Lipschitz.

We hope that this note offers the basis for the development of novel sampling approaches to several problems that fit into the union of subspaces framework.
On the one hand, we have shown on several examples, how bi-Lipschitz sampling operators can be constructed. On the other hand, we have suggested an algorithmic framework which can reconstruct signals with near optimal accuracy. Whilst our contribution was theoretical in nature, our results point the way toward practical strategies that can be developed further in order to tackle a given sampling problem. To achieve this, four problems need to be addressed, 1) defining constraint sets $\A$ that capture relevant prior knowledge, 2) designing realisable sampling operators that satisfy the bi-Lipschitz property, 3) implementing efficient ways to store and manipulate the signals on a computer and 4) developing efficient algorithms to project onto the constraint set.

 -------------------------------------------------------------------------


\end{document}